# Transport evidence for the three-dimensional Dirac semimetal phase in ZrTe$_5$

GuolinZheng[1,2*], Jianwei Lu[1,2*], Xiangde Zhu[1,3], Wei Ning[1,3 §], Yuyan Han[1,3], Hongwei Zhang[1,2], Jinglei Zhang[1,3], Chuanying Xi[1,3], Jiyong Yang[1,3], Haifeng Du[1,3], Kun Yang[4], Yuheng Zhang[1,3,5], and Mingliang Tian[1,3,5 §]

[1]*High Magnetic Field Laboratory, Chinese Academy of Sciences, Hefei 230031, Anhui, People's Republic of China*

[2]*University of Science and Technology of China, Hefei 230026, Anhui, People's Republic of China*

[3]*Hefei Science Center, Chinese Academy of Sciences, Hefei 230031, Anhui, People's Republic of China*

[4]*NationalHigh Magnetic Field Laboratory, Florida State University, Tallahassee, Florida 32306-4005, USA*

[5]*Collaborative Innovation Center of Advanced Microstructures, Nanjing University, Nanjing 210093, the People's Republic of China*

*These authors contributed equally to this work.

§To whom correspondence should be addressed.E-mail:ningwei@hmfl.ac.cn; tianml@hmfl.ac.cn

# Abstract

Topological Dirac semimetal is a newly discovered class of materials which has attracted intense attentions. This material can be viewed as a three-dimensional (3D) analog of graphene and has linear energy dispersion in bulk, leading to a range of exotic transport properties. Here we report direct quantum transport evidence of the 3D Dirac semimetal phase of layered material $ZrTe_5$ by angular dependent magnetoresistance measurements under high magnetic fields up to 31 T. We observed very clear negative longitudinal magnetoresistance induced by chiral anomaly under the condition of the magnetic field aligned only along the current direction. Pronounced Shubnikov-de Hass (SdH) quantum oscillations in both longitudinal magnetoresistance and transverse Hall resistance were observed, revealing anisotropic light cyclotron masses and high mobility of the system. In particular, a nontrivial π-Berry phase in the SdH oscillations gives clear evidence for 3D Dirac semimetal phase. Furthermore, we observed clear Landau level splitting under high magnetic field, suggesting possible splitting of the Dirac point into Weyl points due to broken time reversal symmetry. Our results indicate that $ZrTe_5$ is an ideal platform to study 3D massless Dirac and Weyl fermions in a layered compound.

## I. INTRODUCTION

Three-dimensional (3D) Dirac semimetal is a new kind of topological material with a linear energy dispersion in bulk, often viewed as a "3D graphene"[1,2,3,4,5]. The Dirac point can be viewed as a pair of Weyl points coinciding in momentum space, protected by time reversal, inversion symmetries as well as crystalline point-group symmety[3,4,5,6]. The Dirac point can split by breaking either the time reversal or inversion symmetry[7,8], resulting in transformation of a Dirac semimetal to a Weyl semimetal. In this material, exotic quantum phenomena can be observed, such as chiral magnetic effect[9,10,11,12,13,14,15,16], high bulk carrier mobility[17,18,19] and large nonsaturating magnetoresistance[17,18]. These transport properties ignite an extensive interest in both condensed matter physics and potential application in information technologies[20]. A number of materials, such as $Cd_3As_2$[21,22,23,24,25], $Na_3Bi$[26,27], $Bi_{1-x}Sb_x$[9], TlBiSSe[28] and $YbMnBi_2$[29], have been confirmed to be Dirac semimetals, and very recently, a few monopnictides TaAs[30,31,32], NbAs[33], TaP[34] and NbP[35] have also been experimentally suggested to be the Weyl semimetals generated by the broken space inversion symmetry.

Layered material $ZrTe_5$ has been known for decades due to its large thermoelectric power[36,37], resistivity anomaly[38] and quantum oscillations[39,40]. Earlier experimental results exhibited an extremely small ellipsoidal Fermi surface[40] with small effective mass in the chain direction. Recent theoretical work predicts that the $ZrTe_5$ compound is located close to the phase boundary between weak and strong topological insulators[41]. On the other hand, ARPES experiment and transport observation of chiral anomaly effect demonstrated that it is a promising candidate material of 3D Dirac semimetal[10]. Meanwhile, magneto-infrared spectroscopy studies also revealed several hallmarks of 3D Dirac semimetal of $ZrTe_5$, such as line energy dependence of optical conductivity and Landau level splitting[42,43]. Since $ZrTe_5$ is a quasi-two-dimensional (2D) layered material with low symmetry, the appearance of 3D Dirac semimetal is extremely surprising. To date, there is no direct transport evidence, such as the observation of the nontrivial Berry phase, that would allow one to unambiguously identify ZrTe$_5$ as a 3D Dirac semimetal.

Here we present the magnetotransport studies of ZrTe$_5$ single crystalline nanoribbons with different thickness exfoliated from bulk crystals. We observed negative magnetoresistance in different current directions due to chiral anomaly when the magnetic field is parallel to the current, and prominent Shubnikov-de Hass (SdH) oscillations in both longitudinal resistance $R_{xx}$ and the transverse Hall resistance $R_{xy}$. By tracking the SdH spectra, a nontrivial π Berry phase in bulk is obtained which indicates the existence of Dirac Fermions in ZrTe$_5$. Highly anisotropic light effective masses and high Fermi velocity were observed from the angle-dependent SdH oscillations, clearly revealing the 3D nature of the Fermi surface and strongly indicating a 3D (instead of 2D) Dirac semimetal phase in layered material ZrTe$_5$. Furthermore, we observed clear Landau level splitting with magnetic field approaching 10 T, which provides evidence of splitting the Dirac node into two Weyl nodes in ZrTe$_5$ by breaking the time reversal symmetry. Our experiments provide clear transport evidences for 3D Dirac semimetal phase in layered material ZrTe$_5$.

## II. RESULTS AND DISCUSSION

ZrTe$_5$ single crystals were grown via the iodine vapor transport method in a two-zone furnace with elements Zr (99.99%) and Te (99.99%) as described in Ref. 40. The obtained crystals show a thin elongated rectangular shape with length up to several centimeters. It consists of the prismatic ZrTe$_6$ chains along the crystallographic *a*-axis linked along the *c*-axis via zigzag chains of Te atoms to form a two-dimensional (2D) layer, which stacks along the *b*-axis into a crystal[44]. The crystal structures were confirmed by x-ray diffraction measurements with no trace of impurity phases. Here the ZrTe$_5$ nanoribbons with different thickness were obtained by mechanically exfoliating the bulk crystal. The Hall bar structures were defined by standard electron-beam lithography followed by Au (80 nm)/Ti(10 nm) evaporation and lift-off process. An image of a Hall bar device with thickness of about 80 nm is shown in the inset of Fig.1. Low-field transport measurements were carried out with the physical properties measurement system (PPMS), equipped with a 16 T superconducting magnet. Measurements at high magnetic field were performed using standard ac lock-in techniques with a He-3 cryostat and a dc-resistive magnet (~35 T) at the China High Magnetic Field Laboratory (CHMFL) in Hefei. Magnetotransport experiments were performed with current applied along the long axis (*a*-axis) of the ribbon. Figure 1 shows the resistance as a function of temperature for two samples S1 and S2, with thickness of 160 nm and 80 nm respectively. As temperature decreases, the resistance shows a broad peak at about 145 K, which is known as the "resistivity anomaly", consistent with earlier experiments[38,39,40]. We also noticed that a few recent work has reported the "resistivity anomaly", at a much lower temperatures around 100 K[10,42]. Such a difference may be due to the variations of the crystal quality made with different processes.

Figure 2(a) shows the angular-dependent longitudinal magnetoresistance (LMR), $R_{xx}(B)/R_{xx}(0)$, of sample S1 (160 nm) measured at 2 K by rotating the magnetic field (*B*) in the *a-b* plane, where the angle (*θ*) is tilted from *b*-axis to *a*-axis and the current (*I*) is injected along the *a*-axis. To eliminate the possible influence of the Hall signal due to the asymmetry of the electrodes, the $R_{xx}$ data were averaged by measuring the resistivity over positive and negative field directions. At *θ*=0°, *i.e.*, the magnetic field is applied perpendicular to the *ac*-plane and current (*I*), the LMR is positive at low fields and presents a quadratic classical behavior. With increasing *θ*, the LMR decreases considerably. When the magnetic field is parallel the electric current (i.e., *θ* =90°), we observed a clear negative LMR as shown in the inset of Fig.2(a). In the low field range (*B*< 1 T), the LMR actually shows a positive dip, which may come from the weak antilocalization (WAL) effect stemming from the strong spin-orbit interactions[45]. When the field is increased, the LMR turns negative and then positive at *B* > 4 T. We noticed that the negative LMR is very sensitive to the direction of the magnetic field and disappears when the field is about 1.2° away from the current direction as shown in Fig.2(b). Meanwhile, the negative LMR is suppressed by increasing temperature and disappears above 20 K as shown in Fig.2(c). Such a negative LMR behavior have been observed in TaAs,[11] NbAs,[16] and Na$_3$Bi.[14] It was regarded as one of the most prominent signatures in transport for the chiral anomaly associated to the Dirac and Weyl semimetals. We have repeated the measurement *in situ* with *I* applied along the *c*-axis, where the experiment was made with a quasi-four-probe measurement by injecting current between the contacts 2, 4 or 3, 5 as shown in Fig.1(b). The results are shown in Fig.2(d)-(f). In the high-field range, the LMR behavior is similar to that with current along *a*-axis. At the low-field range, the LMR shows negative behavior below 0.75 T and then positive with larger fields. It disappears when the field is about 5° away from the current direction or the temperature is above 50 K, as shown in Fig.2(d) -2(f).Theoretically, the negative LMR induced by the chiral magnetic effect (CME) shows a quadratic field dependence[9-11]. We have fitted the LMR at low field by the semi-classic formula $\sigma(B) = (1 + C_W \cdot B^2) \cdot \sigma_{WAL} + \frac{1}{\rho_0 + A \cdot B^2}$, where $\sigma_{WAL} = (a \cdot \sqrt{B} + \sigma_0)$ is regarded as the weak antilocalization term. Parameter A and $a$ are fitting parameter and $C_w$ is a positive constant, $\sigma_0$ is the zero field conductivity, and $\rho_0$ is the zero field resistivity. Our results suggested that all CME conductivity are consistent with the quadratic field dependence at low field range, which indicates that the observed highly angle-sensitive negative LMR is most probably due to the chiral magnetic effect in topological semimetal. For Dirac semimetals the CME exists for any direction of the current, while for Weyl semimetal the current direction needs to have a projection along the vector connecting the two isolated Weyl nodes. The negative LMR observed exactly under the condition of *B* parallel to the current injected along different directions indicates that it is a very robust behavior in ZrTe$_5$, and strongly suggests that it hosts the signature of Dirac semimetals.

In addition to the negative LMR in the low-*B*-range, clear resistance oscillations can be observed in the high-field range. To uncover the nature of the quantum oscillations, and get further information of the charge properties of ZrTe$_5$, we have measured $R_{xx}$ and the Hall resistance $R_{xy}$ at different temperatures with a field applied along the *b*-axis (i.e., $B \perp ac$-plane). Figure 3 (a) shows the Hall resistance $R_{xy}$ of sample S2 (80 nm). The $R_{xy}$ shows a negative slope in the low field range (<4 T) and then turns positive in the high field range (>4T). This is because the electron carriers dominated in low fields are almost in quantum limit at magnetic fields above 4 T; thus the hole carriers with relative massive mass become dominant in the transport. From the steep slope near zero field, the electron carrier mobility, $\mu = R_H/\rho_{xx}$ and the carrier density, *n*, are respectively estimated to be $4.6 \times 10^4$ cm$^2$/Vs and $6.1 \times 10^{17} cm^{-3}$ at 2 K, where $\rho_{xx}$ is the longitudinal resistivity and $R_H$ is the Hall coefficient defined by $1/ne$. These results are consistent with the previous reports[39].

By subtracting a smooth background of $R_{xy}$ below 4T, the variations of $\Delta R_{xy}$ versus $1/B$ is shown in Fig.3(b). $\Delta R_{xy}$ oscillates periodically with $1/B$, and the oscillation amplitudes decrease gradually as *T* increases from 2 K to 20 K, indicating that these oscillations originate from the magnetic field-induced quantization of the Landau energy levels, *i.e.*, the SdH oscillations. A single periodic oscillation frequency $F=1/\Delta\left(\frac{1}{B}\right)$ is determined to be 3.76 T with *B* along *b*-axis by Fast Fourier Transform (FFT). According to the Onsager relation, $F = (\hbar/2\pi e)S_F$, we obtain an extremely small cross section of Fermi surface $S_F = 3.8 \times 10^{-4}$Å$^{-2}$, and the Fermi wave vector $k_F \approx 0.011$ Å$^{-1}$. Such a small Fermi cross-section hints that the Fermi level is very close to the Dirac point. Using the Lifshitz-Kosevich formula[46], the SdH oscillation amplitude can be written as

$$\frac{\Delta R(T,B)}{R(B=0)} \propto \left(\frac{\hbar\omega_c}{E_F}\right)^{\frac{1}{2}} \frac{2\pi^2 k_B T/\hbar\omega_c}{\sinh[2\pi^2 k_B T/\hbar\omega_c]} \exp\left(-2\pi^2 k_B T_D/\hbar\omega_c\right). \qquad (1)$$

Where $T_D$ is the Dingle temperature, $\hbar$ is Plank's constant and $k_B$ is the Boltzmann constant. The cyclotron frequency $\omega_c = eB/m^*$, with $m^*$ the effective cyclotron mass at the Fermi energy, which can be estimated by $E_F = m^* v_F^2$, with $v_F$ the Fermi velocity. Figure 3(c) shows the temperature dependence of the relative oscillatory component $\Delta R_{xy}/\Delta R_{xy}(2K)$ for the second Landau level. The solid line is the fitting curve, the fitting generates a small cyclotron mass $m^* = 0.026\, m_e$, and correspondingly, the Fermi velocity calculated from cyclotron mass is $v_F = \hbar k_F/m^* = 4.89 \times 10^5 m/s$. This value is close to the results obtained from the ARPES results and magneto-optics results[10,42,43]. The estimated Fermi level $E_F$ is about 35.4 meV, which is comparable to the well-studied semimetals[11,14,15].

According to the Lifshitz-Onsager quantization rule $S_F \frac{\hbar}{eB} = 2\pi(n + \frac{1}{2} + \beta + \delta)$, where $2\pi\beta$ is the Berry phase and $2\pi\delta$ is the additional phase shift. $\delta$ is determined by the dimensionality of the Fermi surface and the value changes from 0 for 2D to ±1/8 for 3D case[46]. The nontrivial Berry phase is generally considered to be key evidence for Dirac fermions and has been observed in all Dirac materials such as topological insulator[47], Dirac semimetal $Cd_3As_2$,[24] $SrMnBi_2$,[29] and monopnictide Weyl systems[15,16,18,19]. For a Dirac system, the nontrivial Berry phase would make the intercept nonzero. Figure 3(d) shows the Landau index $n$ versus $1/B$, defined from the peak position in Fig.3(b) by integer indices while the valley position defines the half indices. The best linear fit of the curve generates an intercept of the $n$-axis with the value of $|\gamma| \approx 0.18$, as shown in the inset of Fig.3(d). Similar data were obtained by analyzing the SdH oscillations of another sample S1, where the intercept $|\gamma|$ is about 0.17. The nonzero intercept with the value of about 1/8 is consistent with the expectation for a 3D Dirac semimetal.

Except for the nontrivial Berry phase, the angular dependent quantum oscillation provides another insight on the 3D Fermi surface of the Dirac semimetal. Figure 4(a) and 4(d) show the angular-dependent SdH oscillations in sample S2 with magnetic field tilted in the *a-b* and *b-c* planes at 2K, respectively. The derivatives, $dR_{xy}/dB$ as the function of $1/B_\perp=1/B\cos(\theta)$ or $1/B\cos(\varphi)$ are also shown in Fig.4(b) and (e). It was seen that the positions of the peaks or valleys shift continuously with the increase of the tilted angles, θ or φ, indicating that the SdH oscillations do not come from a 2D Fermi surface but have a 3D character (the observed oscillations in Hall resistance under in-plane conditions might be attributed to a small longitudinal resistivity component). This conclusion was further verified by the angular dependence of the oscillation frequency derived from FFT analysis as shown in Fig.4(c) and 4(f), where the frequencies at different angles deviate from the $1/\cos(\theta)$ or $1/\cos(\varphi)$ relation. These data strongly indicate that ZrTe$_5$ is a 3D Dirac semimetal. We noticed that, in a recent work about ZrTe$_5$[48], quasi-2D Dirac fermions were suggested.

In order to gain further insight on the quantum transport properties of ZrTe$_5$, we have performed magnetotransport measurements with field up to 31 T. Figure 5(a) and 5(b) show the results of longitudinal resistance $R_{xx}$ for sample S2 with magnetic field along the *c*-axis. We found that a magnetic field of 20 T along the *c*-axis can drive the carriers in the *a-b* plane to the first Landau level (quantum limit *n*=1). Meanwhile, it is clearly shown that the first (*n*=1) and second (*n*=2) Landau levels split into two peaks, as indicated by the arrows in Fig.5(a). With increasing temperatures from 0.38 to 20 K, the splitting gradually merges to single peak due to thermal fluctuation, as shown in Fig.5(b). Theoretically, the Landau level splitting can be attributed to Zeeman splitting, which is supposed to be insensitive to the *B*-direction[49,50]. Unfortunately, the angle-dependent transport measurement of the He-3 cryostat on the dc facility of CHMFL is not available at present, the angular dependent measurement in ac-plane at higher fields is required in the future.

To obtain more information of the SdH oscillation in different directions, we have also analyzed the SdH oscillation with magnetic field applied along both the *a*-axis and *c*-axis. These results are included in Table I, which shows the parameters obtained from the SdH oscillations for three planes: *b-c*, *a-c* and *a-b*. The smallest effective cyclotron mass, and largest Fermi velocity can be found in the *a-c* plane. The quantum lifetime and mean free path in the *a-c* plane are also much larger than in the other two planes. These results indicate that the Fermi velocity along *b*-axis is much smaller than the other two directions and consistent with the crystal structure where the *a-c* plane layers stacked along *b*-axis. Additionally, the different results shown in both the *a-b* plane and *b-c* plane also demonstrate a different Fermi velocity and cyclotron mass along the *a*-axis and *c*-axis, and suggest an anisotropy in the *a-c* plane.

We note that, in our ZrTe$_5$ samples, a low magnetic field of 3 T along the *b*-axis can drive the carriers in the *a-c* plane into the quantum limit (*n*=1). Even for the *a-b* plane with larger cyclotron masses, a relative low field of 20 T is enough to reach the quantum limit. These values are much smaller than the one needed in Cd$_3$As$_2$, where a magnetic field up to 43 T is required to drive the samples to reach the quantum limit.[51] This observation suggests that ZrTe$_5$ is a promising candidate material to explore quantum phenomena beyond the quantum limit. Furthermore, in our studies, the Landau level splitting can be observed when the field is approaching 10 T, which is similar to NbP[17] but much lower than that in Cd$_3$As$_2$ and TaP, where magnetic fields up to 20 T and 25 T are required for the LL splitting respectively[49,51]. These observations in ZrTe$_5$ indicate that a ZrTe$_5$ is ideal platform to study 3D massless Dirac Fermions, and the quantum transport properties in the quantum limit.

Table I: The parameters obtained from the SdH oscillation for carriers in three plane *b-c*, *a-c*, and *a-b*.

|  | $F$ (T) | $k_F$ (Å$^{-1}$) | $m_{eff}$ (m$_e$) | $v_F$ (10$^5$ m/s) | $\tau$ (10$^{-13}$ s) | $l$ (nm) | $\mu_{Hall}$ (10$^4$ cm$^2$/Vs) | $\mu_{SdH}$ (10$^4$ cm$^2$/Vs) |
|---|---|---|---|---|---|---|---|---|
| *b-c plane* | 33.4 | 0.032 | 0.19 | 1.94 | 1.62 | 34.1 | \ | 0.15 |
| *a-c plane* | 3.76 | 0.011 | 0.026 | 4.89 | 3.13 | 153.1 | 4.6 | 2.12 |
| *a-b plane* | 25.12 | 0.028 | 0.08 | 4.03 | 0.76 | 27.4 | \ | 0.17 |

### III. CONCLUSION

In conclusion, we have studied the quantum transport properties of high-quality ZrTe$_5$ single crystals with magnetic field up to 31 T. When the magnetic field is parallel to the current, we observed chiral anomaly effect-induced negative magnetoresistance in different current directions. By tracking the SdH oscillations, we obtained a nontrivial π-Berry phase, a small effective electron mass about $m^*=0.03m_e$, and a higher mobility $\mu = 4.6 \times 10^4$ cm$^2$/Vs. The angle-dependent SdH oscillation indicates a 3D Fermi surface in ZrTe$_5$. We found that a low magnetic field (~3 T) along the *b*-axis can drive the carriers in the *a-c* plane reach the quantum limit. Furthermore, we observed a clear Landau level splitting at a relative weak magnetic field. These results give a clear evidence for the existence of the 3D Dirac semimetal phase in layered material ZrTe$_5$, which is also an important system for the research of Weyl fermions in a Dirac semimetal system by breaking the time-reversal symmetry.

**ACKNOWLEDGEMENTS**

The author thanks Professor Anton Burkov, Professor Xi Dai, Professor Dong Qian, Professor Yuanbo Zhang and Professor Haizhou Lu for fruitful discussions. This work was supported by the Natural Science Foundation of China (Grant No.11174294, No.11574320, No.11374302, No.11204312, No.U1432251 and No.U1332139), National Science Foundation of USA (Grant No.DMR-1442366), and the program of Users with Excellence, the Hefei Science Center of CAS and the CAS/SAFEA international partnership program for creative research teams of China.

**Figure Captions**

**FIG. 1. (Color online)** Temperature dependence of the longitudinal resistance $R_{xx}$ for samples S1 (160 nm) and S2 (80 nm); the inset shows the SEM image of a Hall bar device used in the measurement. The out-of-plane direction is *b*-axis while the current is applied along the *a*-axis direction (long axis of the ribbon).

**FIG. 2. (Color online)** (a, d) are, respectively, the angular dependence of longitudinal magnetoresistance (LMR), $R_{xx}(B)/R_{xx}(0)$, of sample S1 (thickness: 160 nm) by tilting the magnetic field B in *ab* and *bc* planes measured at 2 K, where a dc current is injected along the a-axis (i.e., the length direction of the ribbon) in (a) and c-axis (i.e., the transverse direction) in (d), respectively. The insets are the blowup of the plot in the low field range. Negative LMR is found when the magnetic field is aligned parallel to the dc current at $\theta=90°$ or $\varphi=90°$). (b,e) are the $R_{xx}(B)/R_{xx}(0)$ versus *B* at several angles near $\theta=90°$ or $\varphi=90°$, respectively. The negative LMR disappears when the tilted angle $\theta$ or $\varphi$) is off the current directions. (c, f) are the $R_{xx}(B)/R_{xx}(0)$ versus *B* with *B//I* (i.e., along a- and c-axis) at different temperatures. The negative LMR is suppressed by increasing temperatures.

**FIG. 3. (Color online)** (a) The hall resistance, $R_{xy}$, versus *B* aligned along *b*-axis at different temperatures. It presents a negative slope in *B*<4 T and then a sign reversal above 4 T. All curves are offset vertically for clarity. (b) The oscillatory component, $\Delta R_{xy}$ versus $1/B$ under different temperatures, the periodic oscillating behavior with $1/B$ indicates the nature of Shubnikov-de Haas (SdH) oscillations due to Landau quantization of the energy levels. The Landau index number, *n*, is marked with the peak position defined by integer indices. (c) The amplitudes of the oscillatory component $\Delta R_{xy}/\Delta R_{xy}(2K)$ at *n* =2 peak under different temperatures. The red line is the theoretical fit, generating an effective mass $m^* = 0.026\ m_e$. (d) The Landau index *n* versus $1/B$, the best linear fit generates a nonzero intercept about -0.18 with the *n*-axis.

**FIG. 4. (Color online)** (a) and (d) are respectively the $R_{xy}$ versus *B* tilted within *a-b* and *b-c* planes for sample S2.

(b) and (e) are the $dR_{xy}/dB$ as the function of $1/B_\perp$ with $B_\perp = B\cos\theta$ or $B\cos\varphi$, respectively. The arrows indicate the shift for Landau index $n=3$ peak. All curves are offset vertically for clarity. (c) and (f) are the angular dependence of the oscillation frequency derived from FFT analysis from (b) and (e). The FFT frequency deviated from $1/\cos(\theta)$ and $1/\cos(\varphi)$ relation with B tilted in *a-b* and *b-c* planes respectively.

**FIG. 5. (Color online)** (a) The $R_{xx}$ versus B up to 30 T aligned along *c*-axis at different temperatures. (b) The oscillatory component $\Delta R_{xx}$ versus 1/B at different temperatures. The splitting of the peaks of $n=1$ and 2 was clearly seen, as marked by the arrows. The splitting is smeared out with the increase of temperature above 15 K due to thermal fluctuation.

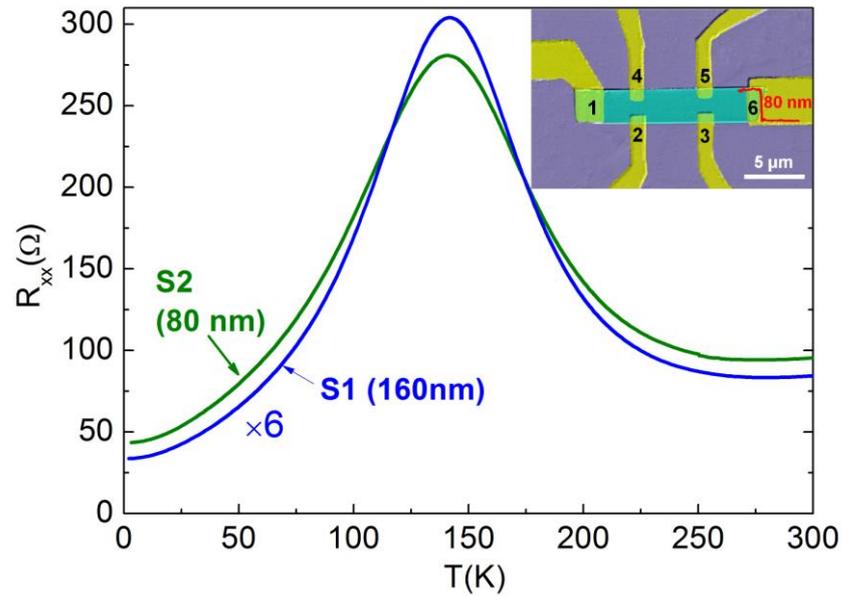

**FIG. 1.**

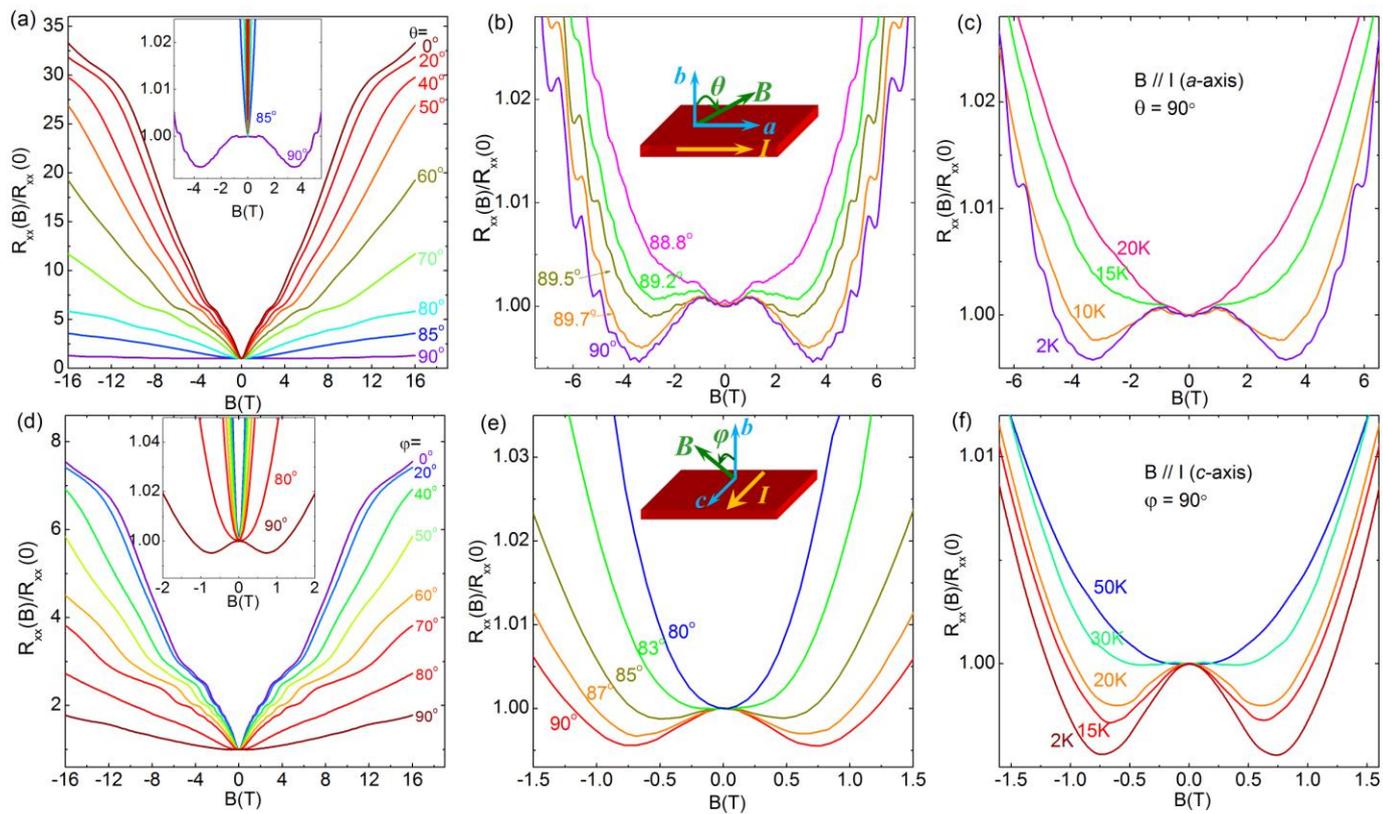

**FIG. 2.**

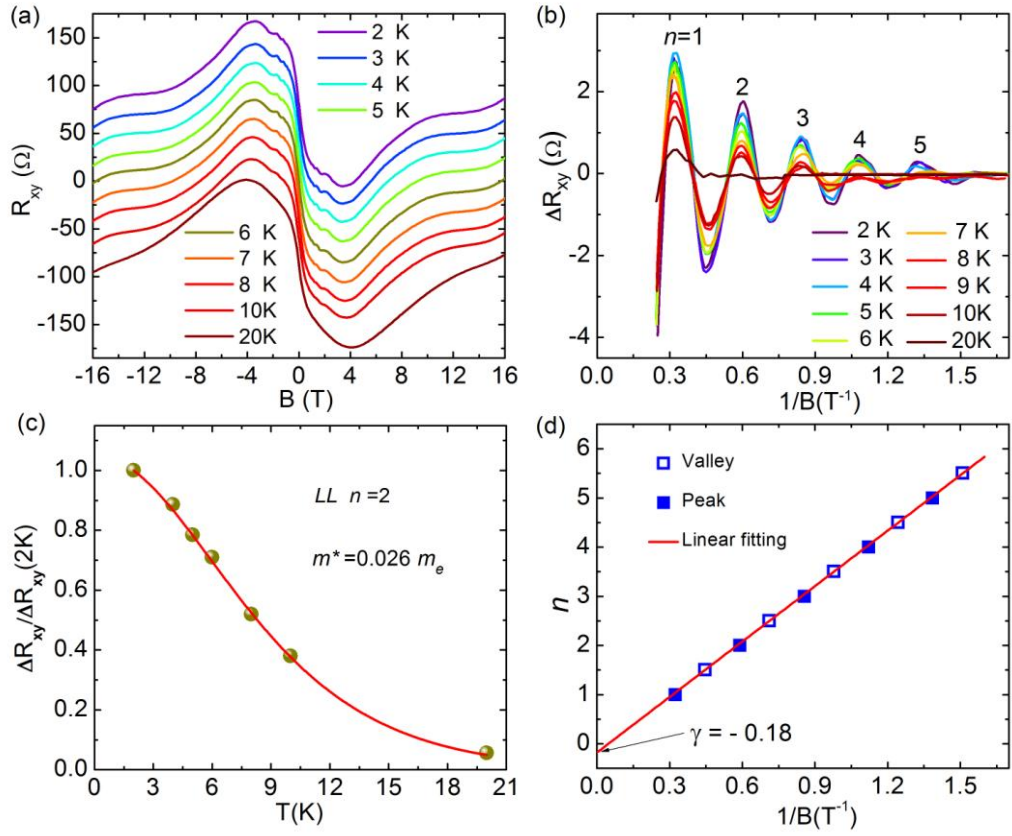

**FIG. 3.**

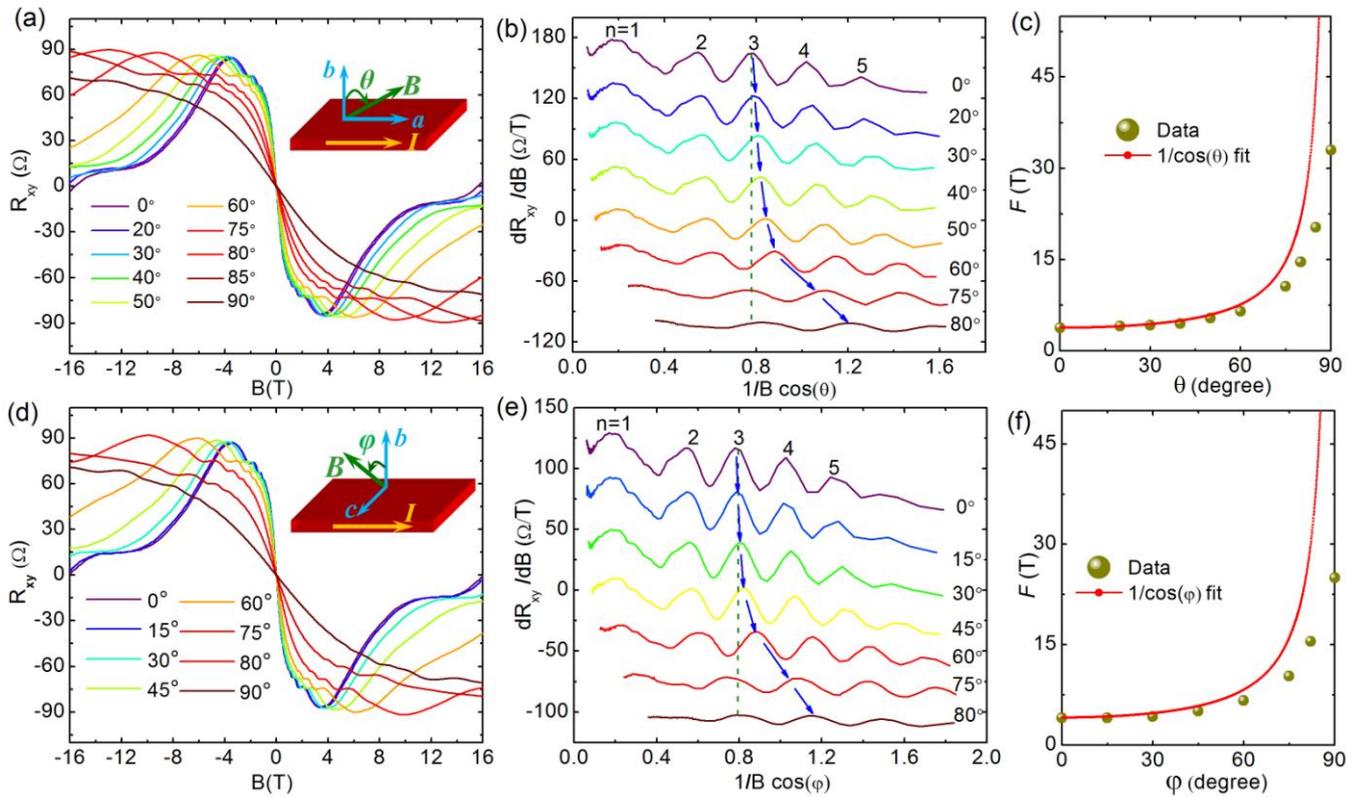

**FIG. 4.**

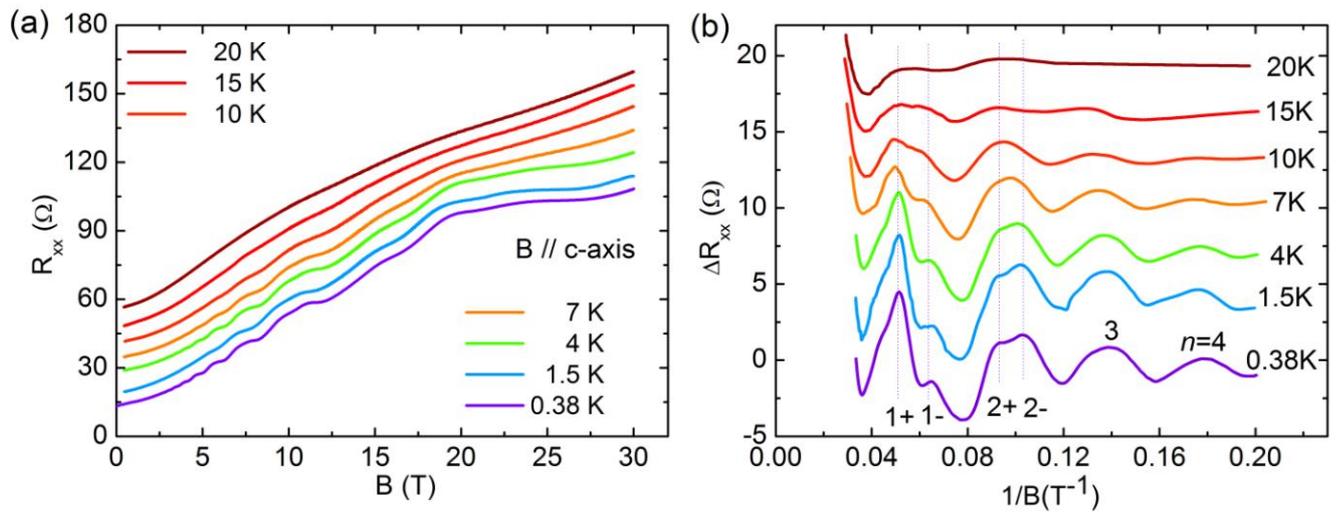

**FIG. 5.**